\documentstyle[epsfig]{mn}
\begin{document}

\title
[Millimetre/submillimetre-wave confusion limits] 
{Observational limits to source confusion in the millimetre/submillimetre 
waveband} 
\author
[A.\,W. Blain, R.\,J. Ivison \& Ian Smail]
{
A.\,W. Blain,$^1$ R.\,J.~Ivison$^2$ and Ian Smail$^3$\\
$^1$ Cavendish Laboratory, Madingley Road, Cambridge, CB3 OHE.\\
$^2$ Institute for Astronomy, Department of Physics \& Astronomy, University 
of Edinburgh, Blackford Hill, Edinburgh, EH9 3HJ.\\
$^3$ Department of Physics, University of Durham, South Road, Durham, 
DH1 3LE.
}
\maketitle

\begin{abstract}
The first observations to detect a population of distant galaxies directly in the 
submillimetre waveband have recently been made using the new 
Submillimetre Common-User Bolometer Array (SCUBA) on the James Clerk 
Maxwell Telescope (JCMT). The results indicate that a large number of distant 
galaxies are radiating strongly in this waveband. Here we discuss their 
significance for source confusion in future millimetre/submillimetre-wave 
observations of both distant galaxies and cosmic microwave background
radiation (CMBR) anisotropies. Earlier estimates of such confusion involved 
significant extrapolation of the results of observations of galaxies at low 
redshifts; our new estimates do not, as they are derived from direct 
observations of distant galaxies in the submillimetre waveband. The results 
have important consequences for the design and operation of existing and 
proposed millimetre/submillimetre-wave telescopes: the {\it Planck Surveyor} 
survey will be confusion-limited at frequencies greater than 350\,GHz, even in 
the absence of Galactic dust emission; a 1$\sigma$ confusion noise limit of about
0.44\,mJy\,beam$^{-1}$ is expected for the JCMT/SCUBA at a wavelength of 
850\,$\mu$m; and the subarcsecond resolution of large 
millimetre/submillimetre-wave interferometer arrays will be required in order to 
execute very deep galaxy surveys. 
\end{abstract}  


\section{Introduction}

Smail, Ivison \& Blain (1997, hereafter SIB) recently detected a new population of 
distant galaxies in a deep survey of gravitational lensing clusters in the
submillimetre waveband using SCUBA on the JCMT (Holland et al. 1998). 
Subsequent deeper integration in these fields, and observations of another five
cluster fields confirm the results (Smail et al. 1998; Blain et al. 1998). The 
identification of the brightest 850-$\mu$m source with a dust-shrouded 
starburst/active galactic nucleus at 
redshift $z=2.8$ (Ivison et al. 1998) and the non-correspondence of the detected 
sources with cluster galaxies support these claims.
Prior to these observations the surface density of faint galaxies in this 
waveband was uncertain by up to three orders of magnitude (Blain \& Longair 
1993, 1996, hereafter BL96), but it is now known to an accuracy of about 50 per 
cent. The estimation of the magnitude of confusion noise (Scheuer 1957) due to 
the submillimetre-wave continuum radiation of discrete extragalactic sources has 
previously required extensive extrapolation from the results of observations of 
low-redshift galaxies (Longair \& Sunyaev 1969; Franceschini et~al. 1989; 
Helou \& Beichman 1990; Fischer \& Lange 1993; Toffolatti et~al. 1995; 
Gawiser \& Smoot 1997), and so the uncertainties in the estimates obtained 
were large as compared with those in the radio and far-infrared wavebands. 
Our new direct observations allow more reliable estimates to be made. 

First, we briefly discuss our observations at 850\,$\mu$m. We introduce a simple  
parametric model of galaxy evolution,  
which is similar in form to the {\it IRAS}-based models discussed by BL96, but 
fitted to the counts derived by SIB, 
to predict the 
properties of confusing sources at other 
millimetre/submillimetre wavelengths. These models include {\it only} the 
contribution from dusty extragalactic objects. At the longest and shortest 
wavelengths in this waveband a significant contribution 
to confusion is expected due to other sources of radiation -- discrete 
extragalactic non-thermal radio emission and Galactic dust emission 
respectively. Secondly, we present our estimates of confusion noise as a 
function of both the angular scale and wavelength of observations. Finally, we 
discuss the relevance of confusion for observations of both distant galaxies and 
the CMBR using existing and planned millimetre/submillimetre-wave telescopes: 
the JCMT/SCUBA; the airborne Stratospheric Observatory For Infrared Astronomy
(SOFIA: Becklin 1997); the Large Millimeter Telescope (LMT/GTM), a 
ground-based 50-m single-antenna telescope (Schloerb 1997) that will
incorporate the BOLOCAM bolometer array receiver (Mauskopf \& Bock 1997); 
large ground-based millimetre/submillimetre-wave interferometer arrays (MIAs: 
Brown 1996; Downes 1996); the ESA {\it Far-Infrared and Submillimetre 
Telescope} ({\it FIRST}: Pilbratt 1997) and {\it Planck Surveyor} (Bersanelli et al. 
1996) space missions, and balloon-borne instruments, for example 
BOOMERANG (Lange et al. 1995). Parameters of these instruments and 
telescopes are listed in Table\,1. An advanced-technology 10-m ground-based 
telescope is also planned for the South Pole (Stark et al. 1998).  

\begin{table}
\caption{Instrumental parameters for instruments and telescopes shown in 
Fig.\,2, and a large MIA. References can be found in Section\,1. It is proposed to 
attach a prototype of the BOLOCAM instrument destined for the LMT to the 
10.4-m Caltech Submillimeter Observatory (CSO). Unless otherwise 
stated the sensitivities quoted refer to a 1$\sigma$ detection in a 1-h
integration including overheads. 
}
{\vskip 0.75mm}
\hrule{\vskip 1.2mm}
\begin{tabular}{ p{2.4cm} p{0.9cm} p{2.0cm} p{2.7cm} } 
Instrument & $\lambda$ / & FWHM Beam- & Sensitivity\\
 & $\mu$m & width / arcmin & \\
\end{tabular}
{\vskip 1.2mm}
\hrule 
{\vskip 1.2mm}
\begin{tabular}{ p{2.4cm} p{0.9cm} p{2.0cm} p{2.7cm} } 
SCUBA         & 450 & 0.12 & 50\,mJy \\
(Jiggle-map) & 850 & 0.22 & 8\,mJy \\
\noalign{\vskip 1.5mm}
{\it FIRST} (25-   & 250 & 0.30 & 3.3\,mJy \\
arcmin$^2$ map)& 350 & 0.42 & 2.9\,mJy \\
                           & 500 & 0.60 & 2.4\,mJy \\
\noalign{\vskip 1.5mm}
{\it Planck Surveyor} & 350 & 4.4 & 26\,mJy \\
(Nominal 14- & 550 & 4.4 & 19\,mJy \\
month mission)	       & 850 & 4.4 & 16\,mJy \\
	       & 1350 & 7.1 & 11\,mJy \\
 	       & 2000 & 10.3 & 11\,mJy \\
\noalign{\vskip 1.5mm}
SOFIA             & 450 & 1.3 & 3.5\,mJy \\
(Point source) & & & \\
\noalign{\vskip 1.5mm}
BOOMERANG & 770 & 20 & $\Delta T/T = $ \\
(Nominal 5-day                       & & & $\>\>3.6\times10^{-6}$ \\
mission) & 1250 & 20 & As above \\
\noalign{\vskip 1.5mm}
BOLOCAM/CSO & 1100 & 0.50 & 0.5\,mJy \\
(36-arcmin$^2$ map) & 1400 & 0.63 & 0.5\,mJy \\
		& 2100 & 0.95 & 0.5\,mJy \\
\noalign{\vskip 1.5mm}
BOLOCAM/LMT & 1100 & 0.10 & 0.033\,mJy \\
(2-arcmin$^2$ map) & 1400 & 0.13 & 0.033\,mJy \\
		& 2100 & 0.20 & 0.033\,mJy \\
\noalign{\vskip 1.5mm}
MIA                & 870 & 0.02 & 0.028\,mJy \\
(single pointing) & 1300 & 0.02 & 0.017\,mJy \\ 
\noalign{\vskip 1.5mm}
\end{tabular}
\hrule
\end{table}

\section{The population of galaxies in the submillimetre waveband}

The galaxies discovered by SIB were detected in the fields of two clusters,
Abell\,370 and Cl2244$-$02, each of which was observed for 25\,ks at 
wavelengths of 450 and 850\,$\mu$m in a circular field 2.3\,arcmin in diameter. 
The 1$\sigma$ noise flux densities in these maps were 14 and 2\,mJy 
respectively. In the total field area of about 10\,arcmin$^2$, six sources were 
detected at 850\,$\mu$m and one at 450\,$\mu$m. The results presented in this
paper are based on these published observations; however, we have now 
expanded our survey to cover five additional clusters, and increased the total 
field area to more than 31\,arcmin$^2$ (Smail et al. 1998; Blain et al. 1998). The 
magnification bias due to gravitational lensing in the clusters (Blain 1997b) was 
exploited by SIB in order to assist the detection of background galaxies; 
however, these results should soon be confirmed by blank-field SCUBA surveys 
(BL96; Pearson \& Rowan-Robinson 1996). After correcting for the moderate 
magnification bias we estimate that the surface density of galaxies with flux 
densities larger than 4\,mJy 
is $(2.5 \pm 1.4) \times 10^3$\,deg$^{-2}$ at a wavelength of 850\,$\mu$m. This 
surface density is larger than that predicted in the most strongly evolving model 
by BL96, but is nevertheless consistent with both the observed intensity of 
diffuse submillimetre-wave background radiation 
(Puget et al. 1996; Schlegel, Finkbeiner \& Davis 1998; Hauser et al. 1998; 
Fixsen et al. 1998) 
and the abundance of heavy elements at the present epoch. In Fig.\,1 the 
cumulative counts of galaxies inferred from the 850-$\mu$m observations are 
compared with the well-fitting model from SIB, which is also used to predict the 
corresponding counts at the wavelengths of four other atmospheric windows -- 
2800, 1300, 450 and 350\,$\mu$m -- and at a wavelength of 175\,$\mu$m, which 
is inaccessible to ground-based telescopes. 
This model is described in SIB, and assumes that the local population of 
{\it IRAS} galaxies (Saunders et al. 1990) undergoes pure luminosity evolution 
of the form $(1+z)^3$ out to $z=2.6$, with a fixed evolution factor of 46.7 in the
range $2.6\le z<7$. The counts predicted by this model at 175\,$\mu$m and
2.8\,mm are consistent with recent observations by Kawara et al. (1997) and
Wilner \& Wright (1997): see Fig.\,1. 
The predicted counts are expected to be most 
accurate at the observed wavelength of 850\,$\mu$m.

\begin{figure}
\begin{center}
\epsfig{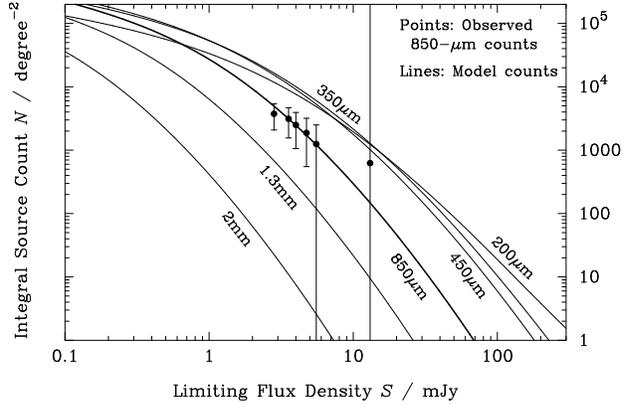}
\end{center}
\caption{Predicted counts of galaxies in the millimetre/submillimetre waveband,
based on the model discussed by SIB. The model is in agreement 
with recent observations by Kawara et al. (1997 -- K) and the 1$\sigma$ upper 
limit of Wilner \& Wright (1997 -- WW). Note that the errors on the SIB counts 
are not independent.}
\end{figure}

\section{Source confusion} 

\begin{figure*}
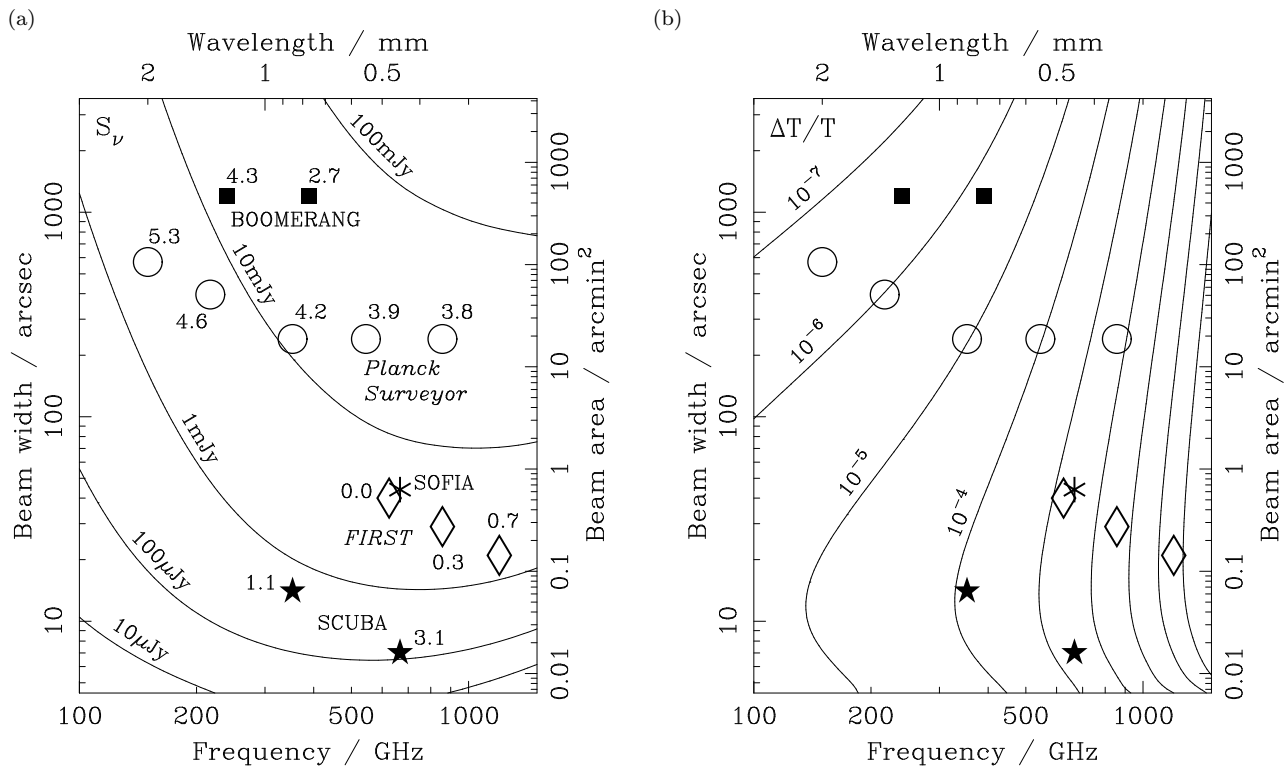

\begin{minipage}{170mm}
(a) \hskip 81mm (b)
\begin{center}
\vskip -5mm
\epsfig{file=figure2a.ps,width=9.8cm,angle=-90} \hskip 9mm
\epsfig{file=figure2b.ps,width=9.8cm,angle=-90}
\end{center}
\caption{The 1$\sigma$ confusion noise predicted using the models shown in 
Fig.\,1 as a function of both observing frequency and angular scale. In (a) the 
noise equivalent flux density $S_\nu$ is plotted. In (b) the relative thermodynamic 
temperature uncertainty in the CMBR spectrum $\Delta T / T$ is plotted. The
contours are spaced by factors of 10, but only the four lowest contours are 
labelled in (b). The resolution limits and observing frequencies of several 
telescopes (Table\,1) are also plotted using different point styles. In (a) the 
two-digit numbers alongside the points represent the logarithms of the 
integration times in hours after which confusion noise is expected to exceed 
instrumental noise for each telescope; for comparison, 1\,day, 
30\,days and 1\,year correspond to values of 1.38, 2.86 and 3.94 respectively. For
BOLOCAM, BOOMERANG, {\it FIRST}, SOFIA and SCUBA these times refer to an 
imaging observation in a single pointing. The ratio of the integration times for 
imaging and
point source photometry assumed for SOFIA is 20, consistent with the estimates
for the {\it FIRST} bolometer camera (Griffin 1997). The time for {\it Planck 
Surveyor} refers to the noise per pixel achieved in an all-sky survey. The 
resolution limit of a large ground-based MIA would lie below the bottom of 
both panels. The corresponding two-digit numbers for an MIA are 0.9 at 
353\,GHz and 1.9 at 230\,GHz, assuming a 3-arcsec beam.}
\end{minipage} 
\end{figure*}

Source confusion from faint galaxies that are not individually recognized can be
calculated rigorously using the formalism derived by Scheuer (1957). However, 
for reasonable source count models confusion is expected
to be most significant when the cumulative surface density of sources $N$ is 
approximately 1\,beam$^{-1}$; the flux density $S$ at this surface density is 
roughly equal to the 1$\sigma$ confusion noise. At larger surface densities many 
sources are expected within each beam, and so the point-to-point fluctuations 
in the detected signal due to the absence or presence of individual sources in 
the beam is reduced. At smaller surface densities, a smaller number of beams 
contain a source, and so the significance of confusion is again reduced. 

The model responsible for the counts shown in Fig.\,1 is used to obtain 
estimates of confusion noise in the millimetre/submillimetre waveband due to 
discrete extragalactic sources of dust continuum radiation. Given that the 
observed counts are currently uncertain by 50 per cent,  
it is reasonable to identify the 1$\sigma$ source confusion noise with the flux 
density at which a count of 1\,beam$^{-1}$ is predicted. A top-hat beam is 
assumed in the calculations, and so other authors can readily adapt the results 
to fit with other beamshapes. Simulations of confusion noise carried out by 
randomly placing sources drawn from the model counts into Gaussian beams
predict values several times larger than the values produced by the 
1\,beam$^{-1}$ analysis. Hence, while this paper does not present the definitive 
calculation of millimetre/submillimetre-wave confusion noise, it is the first
to be observationally-based, makes reliable predictions that represent the 
best state of current knowledge and improves very considerably on the 
important work of earlier authors. 

The resulting estimates of confusion noise as a function 
of both observing wavelength and resolution are presented in 
Fig.\,2. The specified angular resolutions and observing frequencies 
of a range of telescopes and instruments -- the JCMT/SCUBA, {\it FIRST}, 
{\it Planck Surveyor}, BOLOCAM, BOOMERANG and SOFIA -- are also shown 
and listed in Table\,1. The 
integration time required for confusion noise to dominate the noise in an
integration is also given for all the instruments. 
The power-law slopes of the counts in Fig.\,1, from which these estimates are 
determined, are about $\beta = -2$, where $N \propto S^\beta$, and so 
confusion is expected to be insignificant at flux densities about five times larger 
than the confusion noise estimates presented in Fig.\,2(a). 

At frequencies less than about 100\,GHz on angular scales larger than about 
0.2$^\circ$, discrete non-thermal radio sources are expected to start to dominate 
the confusion noise. At the highest frequencies shown in Fig.\,2, Galactic 
emission 
is likely to be the most significant source of confusion noise. The contribution of 
these sources to confusion is not included in Fig.\,2, but extrapolated estimates 
can be found in Franceschini et al. (1989) and Helou \& Beichman (1990). 

The magnification bias owing to lensing is expected to modify the effects of 
confusion in the fields of clusters. If the inner 1-arcmin region of the cluster is 
resolved, then the confusion limits obtained after correcting for the magnification 
bias are expected to be reduced by a factor of about 2 as compared with 
blank fields, because lensing increases both the flux density of background 
galaxies and their separation on the sky. However, if clusters cannot be resolved
on arcminute scales, then the confusion limit is expected to increase by a 
factor of up to about 3. The increase is expected to be largest  
if the observing beam is matched to the Einstein radius of the cluster, and to be 
smaller in larger beams. The effect becomes negligible on scales about 10 times 
larger than the Einstein radius. These effects are discussed in more detail 
elsewhere (Blain 1998a).

\section{Consequences for observations} 

\subsection{Submillimetre-wave blank-field galaxy surveys} 

SCUBA, which operates at wavelengths in the range 2\,mm to 350\,$\mu$m, is 
currently the most sensitive submillimetre-wave instrument on any telescope. 
Blank-field surveys are being carried out using this instrument, and so it is 
important to quantify the flux density at which confusion is expected to 
contribute a significant source of noise to such observations. The 1$\sigma$ 
equivalent flux density of confusion noise in SCUBA observations at wavelengths 
of 450 and 850\,$\mu$m is expected to be about 0.14 and 0.44\,mJy\,beam$^{-1}$ 
respectively. Hence, confusion will impose an effective limit to the depth of a 
blank-field 850-$\mu$m SCUBA survey, but the confusion limit will only be 
reached after about 40 8-hour shifts in a single 2.3-arcmin field, including 
overheads, and so does not present a problem to planned surveys. 

We predict that the 1$\sigma$ confusion limits for the bolometer instrument
on the {\it FIRST} mission (Griffin 1997) are about 2.4, 2.9 and 3.3\,mJy at 
frequencies of 600, 857 and 1200\,GHz respectively. In a single 36-arcmin$^2$ 
field these limits should be reached in integrations lasting about 0.4, 1.2 and 
5.6\,h respectively. Thus relatively shallow large-area galaxy surveys should 
be the natural goals of the {\it FIRST} mission. SOFIA is expected to be confused 
at a 1$\sigma$ flux density of about 5\,mJy at a wavelength of 450\,$\mu$m; 
however, SOFIA is expected to be less sensitive as compared with {\it FIRST} 
(Becklin 1997), and so the confusion limit would only be confronted in a 
considerably longer integration: see Fig.\,2(a). 
Future ground-based interferometer arrays with subarcsecond resolution 
should not suffer from the effects of confusion by extragalactic sources unless 
sensitivities of a few $\mu$Jy are obtained. The prospects for using such arrays 
for extremely deep galaxy surveys are excellent (Blain 1996). Wide-field 
millimetre-wave surveys to a confusion limited depth of order 1 and 0.05\,mJy
will be possible when the BOLOCAM receiver is fitted to the CSO and LMT/GTM 
respectively. The large mapping speed of BOLOCAM ensures that it is an ideal 
instrument to detect sources for subsequent study using the fine angular 
resolution of a MIA. 

\subsection{Observations of the CMBR}

An all-sky millimetre/submillimetre-wave survey of fluctuations in the intensity 
of the CMBR is planned at an angular resolution of about 5\,arcmin and a 
sensitivity $\Delta T/T \sim 10^{-6}$ using the High Frequency Instrument (HFI)
on the {\it Planck Surveyor} mission. The predicted confusion noise in each of the 
five frequency channels of the HFI is presented in Table\,2. These estimates are 
comparable to the intrinsic sensitivity of the HFI in the nominal mission 
(Bersanelli et al. 1996) at the three highest frequencies (353, 545 and 857\,GHz); 
however, confusion noise from dusty galaxies is expected to make only a minor 
contribution to the total noise in the maps at the two lowest observing 
frequencies -- 150 and 217\,GHz. The significant confusion expected at high 
frequencies will be increased by an additional factor due to lensing in the fields
of clusters (Blain 1997a, 1998b), and so {\it Planck Surveyor} may be less 
efficient at 
detecting the Sunyaev--Zel'dovich effect in clusters than suggested by earlier 
studies (Bersanelli et al. 1996; Haehnelt \& Tegmark 1996; Aghanim et al. 1997)

Stratospheric balloon-borne instruments also observe the CMBR in the 
millimetre/submillimetre waveband. The two highest observing frequencies of 
the BOOMERANG instrument, which has a 20-arcmin beam, are 240 and 
390\,GHz. 1$\sigma$ source confusion noise of $\Delta T/T = 3.0\times10^{-7}$ 
and $1.8 \times 10^{-6}$ is expected at these frequencies, respectively, and so 
source confusion noise is expected to be significant only at 390\,GHz in an 
extremely long-duration flight.

There appear to be good prospects for making a ground-based observation of 
CMBR anisotropies on small angular scales using the lowest frequency channel 
of the BOLOCAM receiver at the CSO: see Fig.\,2(b). 1$\sigma$ source confusion 
noise of $\Delta T/T = 3.9\times10^{-6}$ is expected in this channel, at a 
wavelength of 2.1\,mm. Observations of the same field using the two higher 
frequency channels, at wavelengths of 1.4 and 1.1\,mm should allow Galactic 
emission and any bright foreground sources to be subtracted from the 
resulting map. 

Although millimetre/submillimetre-wave observations of the CMBR will be 
complicated by the presence of luminous distant dusty galaxies, these sources 
could be of great interest for both the determination of the star formation rate in 
the early universe (Blain et al. 1998) and the compilation of large samples of 
gravitational lenses (Blain 1996, 1997c, 1998a). 

\begin{table}
\caption{The 1$\sigma$ source confusion noise $(\Delta T / T)_{\rm conf}$ 
expected from dusty galaxies in each of the five observing bands of the 
{\it Planck Surveyor} HFI instrument and the specifications of the instrument, 
including its intrinsic sensitivity $(\Delta T / T)_{\rm sens}$.}
{\vskip 0.75mm}
\hrule
{\vskip 1.2mm}
\begin{tabular}{ p{1.1cm} p{1.1cm} p{1.5cm} p{1.3cm} p{1.3cm} } 
$\lambda$ / mm & $\nu$ / GHz & Pixel Area & 
$(\Delta T / T)_{\rm sens}$ & $(\Delta T / T)_{\rm conf}$ \\
 & & / arcmin$^2$ & & \\
\end{tabular}
{\vskip 1.2mm}
\hrule 
{\vskip 1.2mm}
\begin{tabular}{ p{1.1cm} p{1.1cm} p{1.5cm} p{1.3cm} p{1.3cm} } 
2.00 & 150 & 106 & $1.2 \times 10^{-6}$ & $2.4 \times 10^{-7}$ \\
1.38 & 217 & 51 & $2.0 \times 10^{-6}$ & $1.0 \times 10^{-6}$ \\
0.85 & 353 & 19 & $1.2 \times 10^{-5}$ & $8.1 \times 10^{-6}$ \\
0.55 & 545 & 19 & $7.7 \times 10^{-5}$ & $8.1 \times 10^{-5}$ \\
0.35 & 857 & 19 & $4.2 \times 10^{-3}$ & $4.5 \times 10^{-3}$ \\
\noalign{\vskip 1.5mm}
\end{tabular}
\hrule
\end{table}

\section{Conclusions}

\begin{enumerate}
\item We have estimated the source confusion noise expected in 
millimetre/submillimetre-wave observations, based on direct observations of 
distant galaxies at a wavelength of 850\,$\mu$m (SIB). Although comparable in 
magnitude to previous estimates of the confusion noise, our predictions are
much less dependent on either extrapolations or model-dependent factors.  
\item Reliable limits to the maximum depth of galaxy surveys that can be carried 
out using either ground-based or space-borne telescopes are imposed by these 
estimates. For example, a 1$\sigma$ sensitivity limit of about 0.4\,mJy is 
predicted for the JCMT/SCUBA at 850\,$\mu$m. This limit could be
reached in a fully-sampled map after about 300\,h of integration in a 
single field, and so is unlikely to present a problem to planned surveys. Future 
airborne and space-borne submillimetre-wave telescopes are expected to be 
confusion limited much more 
quickly, and so extremely deep millimetre/submillimetre-wave galaxy 
surveys will require the subarcsecond resolving power of large ground-based 
interferometer arrays. 
\item Our estimates will be useful for planning and operating future 
millimetre/submillimetre-wave telescopes. A {\it FIRST} survey should cover a
large area of sky and not involve integrations longer than a few hours in any 
single-pointed field. Confusion noise arising from 
distant dusty galaxies will dominate 
instrumental noise in the planned all-sky survey using the {\it Planck Surveyor} 
HFI instrument at frequencies greater than 350\,GHz. Confusion noise will be 
more severe in the fields of clusters, reducing the efficiency of {\it Planck 
Surveyor} for the detection of the Sunyaev--Zel'dovich effect. 
\end{enumerate} 

\section*{Acknowledgements}
We thank Sarah Church, Andy Harris, Mike Jones, Anthony 
Lasenby, Malcolm Longair, Richard Saunders and Roberto Terlevich for useful 
discussions, and an anonymous referee for his/her helpful comments. The 
collaboration conducting the observations underlying this work now includes 
Jean-Paul Kneib.
RJI and IRS are supported by PPARC Advanced Fellowships.


\begin{thebibliography}{} 

\bibitem[\protect\citename{bl}%
]{A}
Aghanim N., De Luca A., Bouchet F.\,R., Gispert R., Puget J.\,L., 1997,
A\&A, 325, 9

\bibitem[\protect\citename{bl}%
]{SOFIA}
Becklin E.\,E., 1997, in Wilson A. ed., ESA SP-401,The Far-infrared and
Submillimetre Universe. ESA publications, Noordwijk, p.\,201


\bibitem[\protect\citename{bl}%
]{PLANCK}
Bersanelli M. et al., 1996, SCI(96)3, COBRAS/SAMBA. ESA, Paris

\bibitem[\protect\citename{bl}%
]{B96}
Blain A.\,W., 1996, MNRAS, 283, 1340

\bibitem[\protect\citename{bl}%
]{BL97A}
Blain A.\,W., 1997a, in Holt S.\,S., Mundy J.\,G. eds, Star 
formation: near and far. Am. Inst. Phys., Woodbury NY, p.\,606

\bibitem[\protect\citename{bl}%
]{BL97B}
Blain A.\,W., 1997b, MNRAS, 290, 553

\bibitem[\protect\citename{bl}%
]{BL97C}
Blain A.\,W., 1997c, in Wilson A. ed., The Far-infrared and
Submillimetre Universe. ESA SP-401, ESA publications, Noordwijk, p.\,175 
(astro-ph/9710139)

\bibitem[\protect\citename{bl}%
]{BL97D}
Blain A.\,W., 1998a, MNRAS, in press (astro-ph/9801098)

\bibitem[\protect\citename{bl}%
]{BL97E}
Blain A.\,W., 1998b, MNRAS, in press (astro-ph/9801099)

\bibitem[\protect\citename{bl}%
]{BL93A}
Blain A.\,W., Longair M.\,S., 1993, MNRAS, 264, 509

\bibitem[\protect\citename{bl}%
]{BL96}
Blain A.\,W., Longair M.\,S., 1996, MNRAS, 279, 847 (BL96) 

\bibitem[\protect\citename{bl}%
]{BSIK}
Blain A.\,W., Smail I., Ivison R.\,J., Kneib J.-P., 1998, MNRAS, submitted 
(astro-ph/9806062) 


\bibitem[\protect\citename{bl}%
]{BROWN}
Brown R.\,L., 1996, in Bremer M.\,N., van der Werf P., R\"ottgering~H.\,J.\,A., Carilli
C.\,R. eds, Cold Gas at High Redshift. Kluwer, Dordrecht, p.\,411


\bibitem[\protect\citename{bl}%
]{DOWNES}
Downes D., 1996, in Shaver P. ed., Science with Large 
Millimetre Arrays, Springer, Berlin, p.\,16


\bibitem[\protect\citename{blah}%
]{FL}
Fischer M.\,L., Lange A.\,E., 1993, ApJ, 419, 433


\bibitem[\protect\citename{blah}%
]{FIX2}
Fixsen D.\,J., Dwek E., Mather J.\,C., Bennett C.\,L., Shafer R.\,A., 1998, ApJ, 
submitted (astro-ph/9803021)
 
\bibitem[\protect\citename{blah}%
]{FLDD}
Franceschini A., Toffolatti L., Danese L., De Zotti G., 1989, ApJ, 344, 35

\bibitem[\protect\citename{bl}%
]{GS}
Gawiser E., Smoot G.\,F., 1997, ApJ, 480, L1

\bibitem[\protect\citename{bl}%
]{G}
Griffin M., 1997, in Wilson A. ed., ESA SP-401, The Far-infrared and
Submillimetre Universe. ESA publications, Noordwijk, p.\,31

\bibitem[\protect\citename{bl}%
]{HT}
Haehnelt M.\,G., Tegmark M., 1996, MNRAS, 279, 545

\bibitem[\protect\citename{bl}%
]{Hetal}
Hauser M.\,G. et al., 1998, ApJ, submitted
 
\bibitem[\protect\citename{bl}%
]{HB}
Helou G., Beichman C.\,A., 1990, in Kaldeich\,B. ed.,
From ground-based to space-borne sub-millimetre astronomy.  
ESA vol.\,314, ESA Paris, p.\,117

\bibitem[\protect\citename{bl}%
]{Holl}
Holland W.\,S., Gear W.\,K., Lightfoot J.\,F., Jenness T., Robson E.\,I., 
Cunningham C.\,R., Laidlaw K., 1998, in Phillips T.\,G. ed., Advanced Technology
MMW, Radio and Terahertz telescopes. Proc. SPIE Vol. 3357, SPIE, Bellingham, 
in press

\bibitem[\protect\citename{bl}%
]{I+7}
Ivison R.\,J., Smail I., Le Borgne J.-F., Blain A.\,W., Kneib J.-P.,
B\'ezecourt J., Kerr T.\,H., Davies~J.\,K., 1998, MNRAS, in press
(astro-ph/9712161)

\bibitem[\protect\citename{bl}%
]{K}
Kawara K. et al. 1997, in Wilson A. ed., The Far-infrared and
Submillimetre Universe. ESA SP-401, ESA publications, Noordwijk, p.\,285 (K)



\bibitem[\protect\citename{bl}%
]{L}
Lange A. et al., 1995, Space Science Reviews, 74, 145

\bibitem[\protect\citename{blah}%
]{LS}
Longair M.\,S., Sunyaev R.\,A., 1969, Nat, 223, 719

\bibitem[\protect\citename{blah}%
]{M}
Mauskopf P., Bock J., 1997, \hfill \newline 
http://www-lmt.phast.umass.edu/pub/ins/pres\_cam


\bibitem[\protect\citename{bl}%
]{PRR}
Pearson C., Rowan-Robinson M., 1996, MNRAS, 283, 174

\bibitem[\protect\citename{bl}%
]{P}
Pilbratt G., 1997, in Wilson A. ed., The Far-infrared and
Submillimetre Universe. ESA SP-401, ESA publications, Noordwijk, p.\,7


\bibitem[\protect\citename{blah}%
]{PABB}
Puget J.-L., Abergel A., Bernard J.-P., Boulanger F., Burton W. B., D\'esert
F.-X., Hartmann D., 1996, A\&A, 308, L5


\bibitem[\protect\citename{bl}%
]{IRAS}
Saunders W., Rowan-Robinson M., Lawrence A., Efstathiou G., Kaiser N., 
Ellis R. S., Frenk C.\,S., 1990, MNRAS, 242, 318


\bibitem[\protect\citename{blah}%
]{S57}
Scheuer P.\,A.\,G., 1957, Proc. Cam. Phy. Soc., 53, 764 

\bibitem[\protect\citename{blah}%
]{SFD}
Schlegel D.\,J., Finkbeiner D.\,P., Davis M., 1998, ApJ, 499, in press
(astro-ph/9710327)

\bibitem[\protect\citename{blah}%
]{LMT}
Schloerb F.\,P., 1997, in Latter W.\,B., Radford S.\,J.\,E., Jewell~P.\,R., Mangum
J.\,G., Bally J. eds, 25 years of millimeter spectroscopy. Proc. IAU 170, Kluwer, 
Dordrecht, p.\,221
 
\bibitem[\protect\citename{blah}%
]{SIB}
Smail I., Ivison R.\,J., Blain A.\,W., 1997, ApJ, 490, L5 (astro-ph/9708135; SIB)

\bibitem[\protect\citename{blah}%
]{SIBK}
Smail I., Ivison R.\,J., Blain A.\,W., Kneib J.-P., 1998, ApJL, submitted 
(astro-ph/9806061) 

\bibitem[\protect\citename{blah}%
]{SP}
Stark A.\,A., Carlstrom J.\,E., Israel F.\,P., Menten K.\,M., Peterson~J.\,B., Phillips
T.\,G., Sironi G., Walker W.\,W., 1998, in Phillips T.\,G. ed., Advanced Technology
MMW, Radio and Terahertz telescopes. Proc. SPIE vol. 3357, SPIE, Bellingham, 
in press (astro-ph/9802326)

\bibitem[\protect\citename{blah}%
]{TOF}
Toffolatti L. et al., 1995, Astron. Lett. and Comm., 32, 125

\bibitem[\protect\citename{blah}%
]{WW}
Wilner D.\,J., Wright M.\,C.\,H., 1997, ApJ, 488, L67 (WW)

\end{thebibliography}
\end{document}